\documentclass[sigconf]{acmart}

\usepackage{amsmath,amssymb,amsfonts}
\usepackage{algorithmic}
\usepackage[english]{babel}
\usepackage{algorithm}
\usepackage{enumitem}
\usepackage{graphicx}
\usepackage{textcomp}
\usepackage{makecell}
\usepackage{xcolor}
\usepackage[font=small,labelsep=colon]{caption}
\usepackage{subcaption}
\usepackage{multirow}
\usepackage{libertine}
\AtBeginDocument{%
  }

\copyrightyear{2025}
\acmYear{2025}
\setcopyright{acmlicensed}\acmConference[NOSSDAV '25]{The 35th edition of the Workshop on Network and Operating System Support for Digital Audio and Video}{March 31-April 4 2025}{Stellenbosch, South Africa}
\acmBooktitle{The 35th edition of the Workshop on Network and Operating System Support for Digital Audio and Video (NOSSDAV '25), March 31-April 4 2025, Stellenbosch, South Africa}
\acmDOI{10.1145/3712678.3721886}
\acmISBN{979-8-4007-1469-6/2025/03}
\begin{document}

\title{Video Streaming with \textit{Kairos}: An MPC-Based ABR with Streaming-Aware Throughput Prediction}

% \author{Anonymous NOSSDAV submission}
% \affiliation{Shanghai Jiao Tong University, Shanghai, China}
\author{Ziyu Zhong}
\affiliation{%
    \institution{Shanghai Jiao Tong University}
    \city{Shanghai}
    \country{China}
}
\email{marsink@sjtu.edu.cn}

\author{Mufan Liu}
\affiliation{%
    \institution{Shanghai Jiao Tong University}
    \city{Shanghai}
    \country{China}
}
\email{sudo_evan@sjtu.edu.cn}

\author{Le Yang}
\affiliation{%
    \institution{University of Canterbury}
    \city{Christchurch}
    \country{New Zealand}
}
\email{le.yang@canterbury.ac.nz}

\author{Yifan Wang}
\affiliation{%
    \institution{Shanghai Jiao Tong University}
    \city{Shanghai}
    \country{China}
}
\email{yifanw9@sjtu.edu.cn}

\author{Yiling Xu}
\authornote{Yiling Xu is the corresponding author.}
\affiliation{%
    \institution{Shanghai Jiao Tong University}
    \city{Shanghai}
    \country{China}
}
\email{yl.xu@sjtu.edu.cn}

\author{Jenq-Neng Hwang}
\affiliation{%
    \institution{University of Washington}
    \city{Seattle}
    \country{United States}
}
\email{hwang@uw.edu}
% \author{Ziyu Zhong, Mufan Liu, Le Yang, Yifan Wang, Yiling Xu, Jenq-Neng Hwang}
% \affiliation{%
%     \institution{Shanghai Jiao Tong University}
%     \city{Shanghai}\country{China}
% \email{dave,judy,steve@university.edu}
% \email{firstname.lastname@phillips.org}

\begin{abstract}
Throughput prediction in current adaptive bitrate (ABR) schemes often neglects streaming-aware characteristics, such as sequence irregularity and prediction smoothness, resulting in inaccurate predictions and suboptimal performance. To address these challenges, we propose Kairos, an MPC-based ABR scheme that integrates an attention-based throughput predictor with buffer-aware uncertainty control to enhancing both prediction accuracy and adaptability to dynamic network conditions. Specifically, Kairos employs a multi-time attention network (mTAN) to process irregularly sampled streaming data, producing uniformly spaced latent representations. Based on these, we introduce a percentile prediction network to estimate future throughput percentiles, along with a buffer-aware uncertainty control module that selects the optimal percentile based on the current buffer status. As smoothness is another key component of QoE, we incorporate a smoothness regularizer to ensure consistent throughput predictions, thereby facilitating smoother ABR decisions. Our Kairos design integrates sampling irregularity, prediction uncertainty, and smoothness into the throughput prediction, significantly enhancing bitrate decision making within the MPC framework. Extensive trace-driven and real-world experiments demonstrate that Kairos outperforms state-of-the-art ABR schemes, achieving a QoE improvement ranging from 6.42\% to 29.45\% across diverse network conditions. 
\end{abstract}





\maketitle
% \vspace{-0.6cm}
\section{Introduction}
Online video streaming has transformed the global media landscape and influenced viewing behaviors. Driven by the increasing demand, the global video streaming market revenue reached an estimated \$288 billion in 2023 \cite{Statista}. To enhance the quality of experience (QoE) of users, companies such as YouTube and Netflix have already leveraged adaptive bitrate (ABR) techniques to deliver top-notch viewing services.

ABR is a streaming control algorithm that adaptively switches encoded video representations to enhance user QoE (\cite{Survey1, Survey2, Survey3, PENSIEVE, EVAN, StreamOptix}). It divides the video into small chunks, each encoded at multiple bitrates, and dynamically adjusts the transmission bitrate based on network conditions (e.g., buffer size, throughput) to optimize QoE. Existing ABR research can be categorized into learning-based and model-based methods. Learning-based ABRs (\cite{PENSIEVE, ANT, OLNC, multiuser}) use deep neural networks to learn bitrate decision policies, achieving the state-of-the-art (SOTA) performance in training-like scenarios. However, these methods depend heavily on training samples and lack interpretability, which can lead to suboptimal decisions in unseen cases. On the other hand, model-based methods, also known as "white-box" strategies, improve ABR performance through theoretical insights and can better adapt to a variety of network scenarios compared to learning-based approaches \cite{BayesMPC}. A notable model-based algorithm is model predictive control (MPC) \cite{MPC}, which optimizes bitrate decisions by considering both throughput and buffer occupancy. MPC has shown potential for achieving theoretically optimal performance and is widely used in academia. However, its effectiveness heavily relies on the accuracy of throughput predictions. Unfortunately, throughput prediction is inherently ill-posed, making high accuracy unattainable and often resulting in reduced QoE. To address this challenge, our approach seeks to enhance ABR decisions by increasing the predictive model's capacity. Here, "capacity" encompasses not only prediction accuracy but also the model's ability to adapt to and respond to the unique characteristics of streaming, thereby making the predictor "streaming-aware."

\begin{figure}[t]
    \centering
    % \vspace{-0.4cm}
    
    \begin{subfigure}[b]{0.22\textwidth}
        \centering
        \includegraphics[width=\textwidth]{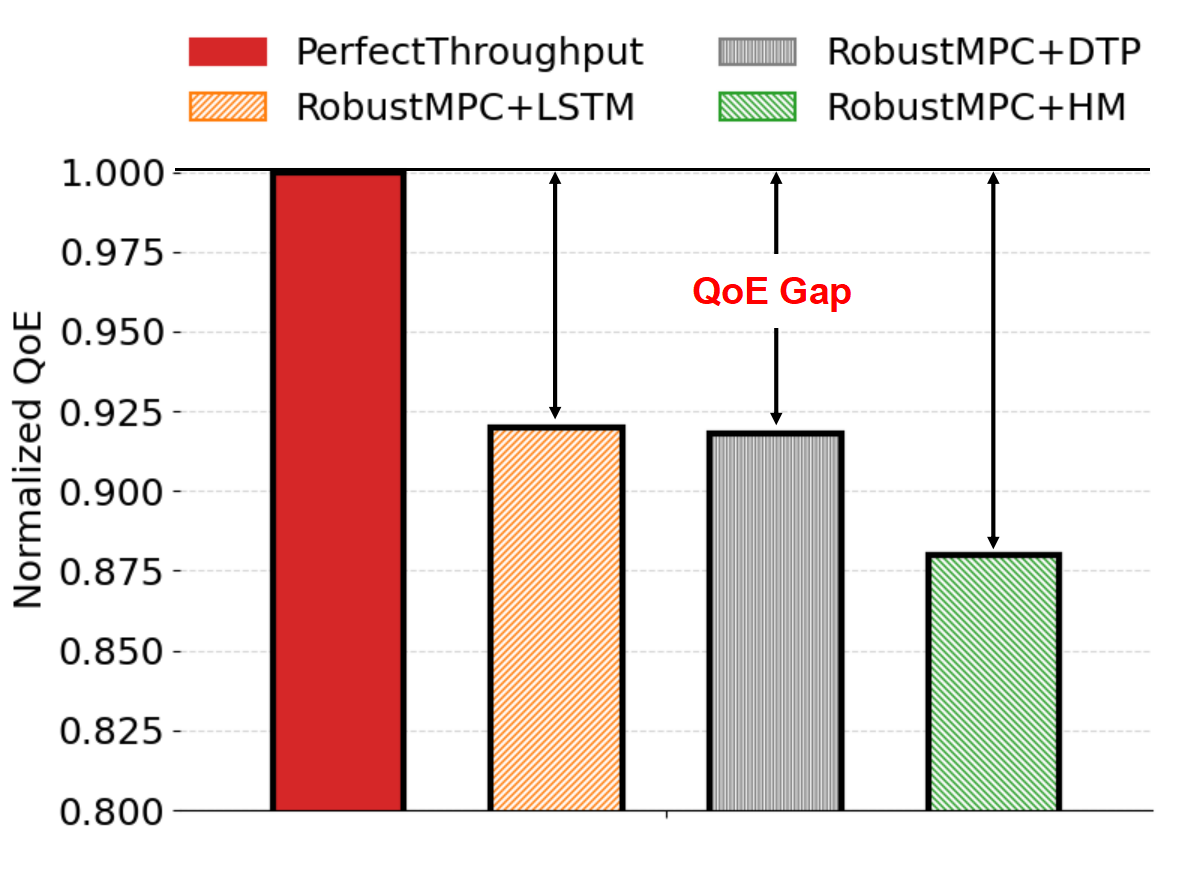}
        % \vspace{-0.6cm}
        \caption{QoE gap}
        \label{fig:image1}
    \end{subfigure}
    \hspace{0.2cm}
    \begin{subfigure}[b]{0.22\textwidth}
        \centering
        \includegraphics[width=\textwidth]{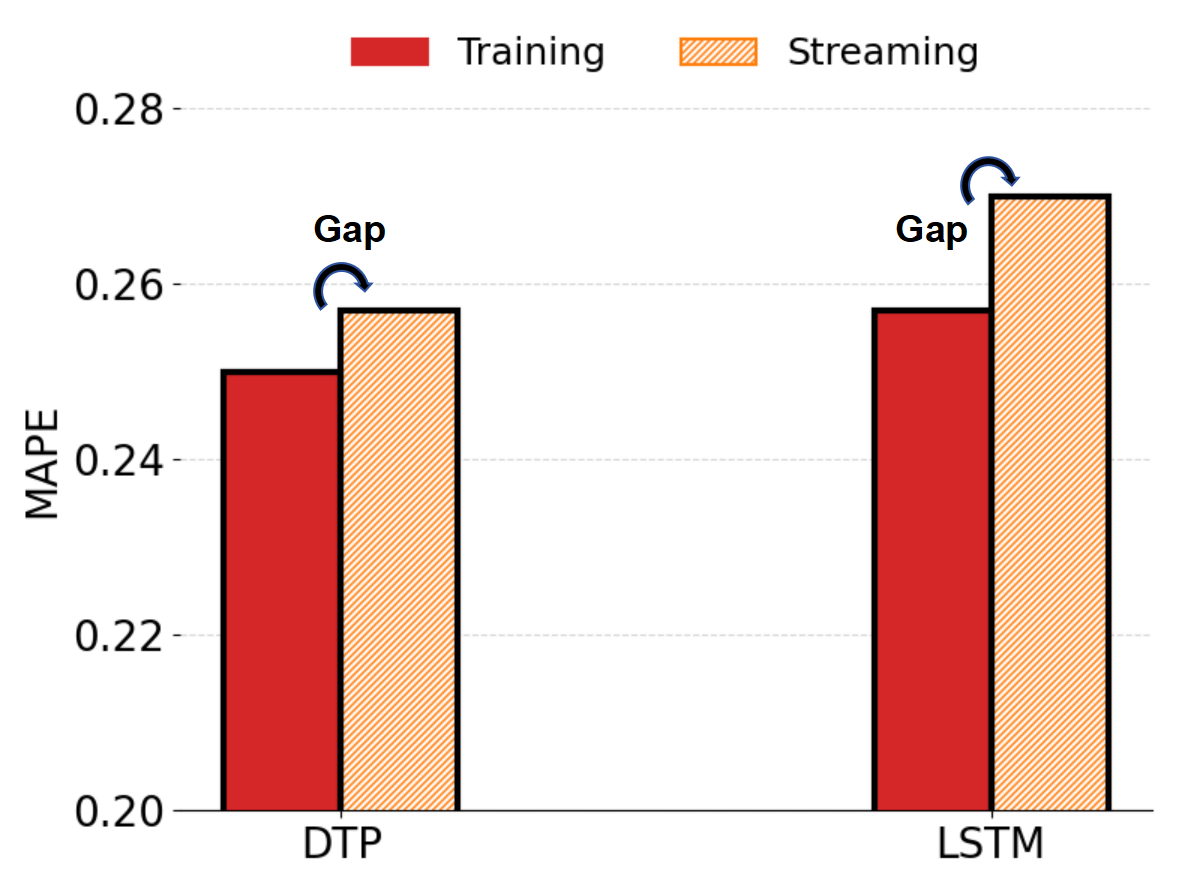}
        % \vspace{-0.6cm}
        \caption{Throughput pred. gap}
        \label{fig:image2}
    \end{subfigure}
    % \vspace{-0.4cm}
    \caption{Performance gap between MPC-based ABRs.}
    \label{fig:both_images}
    % \vspace{-0.6cm}
\end{figure}

\begin{figure*}[t]
\centering 
    % \vspace{-0.2cm}
  \begin{subfigure}[b]{0.6\textwidth}
    \includegraphics[width=\textwidth]{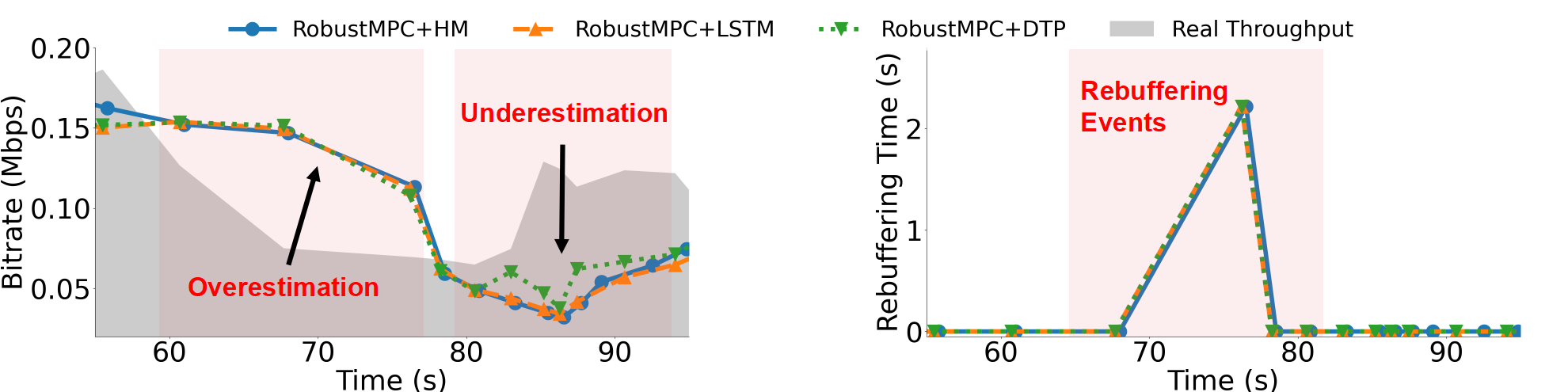}
    % \vspace{-0.5cm}
    \caption{Performance degradation caused by imprecise discount factor of RobustMPC.}
    \label{fig:sub4}
  \end{subfigure}
  \hspace{2mm} % Adjusts the spacing between images
  % \hspace{-5mm} % Adjusts the spacing between images
  \begin{subfigure}[b]{0.3\textwidth}
    \includegraphics[width=\textwidth]{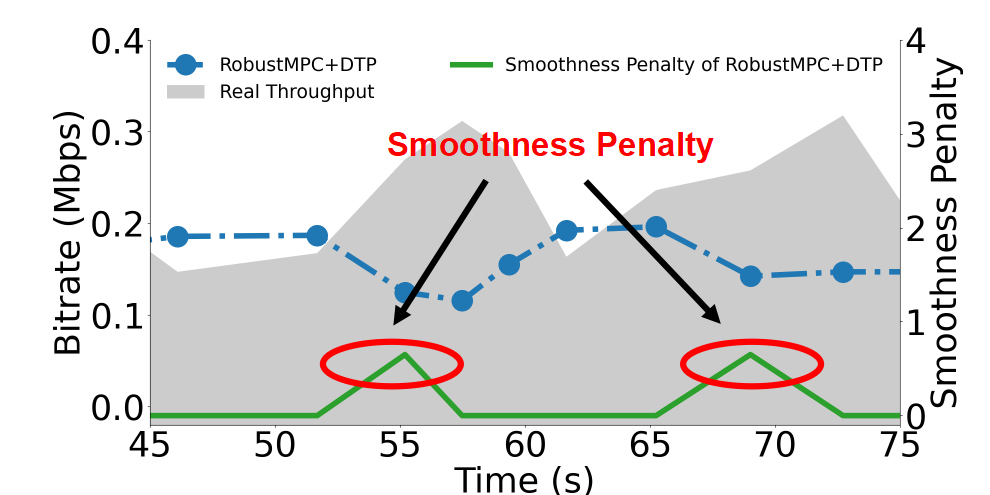}
    % \vspace{-0.5cm}
    \caption{Effect of smoothness penalty.}
    \label{fig:sub5}
  \end{subfigure}
% \vspace{-0.35cm}
\caption{The main bottleneck limiting MPC-based ABR performance is the performance of throughput predictor.}
\label{fig:example-image} % Image label for referencing
% \vspace{-0.4cm}
\end{figure*}

In response to the aforementioned streaming challenges, we propose Kairos, an MPC-based ABR scheme that integrates an attention-based throughput predictor with buffer-aware uncertainty control to enhance both prediction accuracy and adaptability to dynamic network conditions. Specifically, our streaming-aware throughput prediction design \textit{focuses on three key aspects: sequence irregularity, prediction uncertainty, and smoothness}. Existing throughput prediction methods, whether rule-based or learning-based, overlook the fact that the streaming nature leads to irregularly sampled sequences \cite{ISS}, which significantly impacts prediction accuracy. For instance, historical data are collected at the chunk level, but since the download time for each chunk varies, the chunk-level samples are not uniformly distributed across the streaming duration. To address this issue, we adopt a multi-time attention network (mTAN) \cite{mTAN} that interpolates chunk-level samples at regular intervals to reconstruct latent representations uniformly distributed over time. These representations are inherently time-regular, thus improving the accuracy of subsequent predictions. In addition to sequence irregularity, we also identify that overly aggressive or conservative throughput predictions during streaming can result in rebuffering or inefficient network utilization. To mitigate this issue, we propose a novel prediction network that estimates future throughput percentiles and incorporates a dedicated buffer-aware uncertainty control module. This module dynamically adjusts the lower bound of throughput predictions based on buffer occupancy, thereby effectively constraining the sensitivity of the throughput predictions to fluctuating network conditions. Moreover, smoothness is integral to optimizing user QoE, yet existing methods often prioritize it only during the ABR decision-making phase, neglecting its regularization during throughput prediction. By incorporating a smoothness regularizer directly into the training of our throughput predictor, we ensure that predicted throughput values transition more smoothly, thereby enhancing the overall streaming experience and QoE. In summary, Kairos effectively extracts features from irregularly sampled sequences, incorporates streaming uncertainty for more robust bitrate decisions, and integrates smoothness into the predictor’s training to ensure seamless transitions.

Our contributions can be summarized as follows 
% \textcolor{red}{(can be reorganized into 3 points: irregularity; uncertainty+smoothness; experiments)}:
% \begin{itemize}[leftmargin=*]
\begin{itemize}[leftmargin=*]
\item We identify several key factors in jointly optimizing throughput prediction with ABR. Results show that integrating streaming characteristics into throughput prediction can enhance bitrate decisions and achieve higher QoE metrics. 
% \item We proposed 
\item We propose Kairos, a novel ABR scheme that leverages a streaming-aware throughput predictor mTAN-TP with a buffer-aware uncertainty adjustment strategy to generate the lower bound of future throughput, which subsequently helps MPC make better bitrate decisions. 
% Results show that Kairos can effectively reduce the rebuffering and smoothness penalty while preserving higher bitrate utility.
\item We conduct both trace-driven and real-world experiments to validate the performance of Kairos. The results verify that Kairos outperforms existing SOTA approaches by 6.42\% to 29.45\% on QoE metrics, which proves the effectiveness of the proposed scheme.
\end{itemize}

\section{Background, Motivations and Key Ideas}
% \textcolor{red}{(this part can be shortened, for each Q\&A pair, different marks should be added before "our solution" to make it more clear)}:
% \par An ABR algorithm seeks to improve QoE by dynamically selecting chunks of different qualities according to the network condition reported from the client. 
% \par Among the existing ABR techniques, 
\par MPC-based ABRs are popular due to their easy deployment and consistent performance across diverse network conditions while ensuring interpretability. It is found that one main factor limiting the performance of the MPC-based ABR is the throughput prediction \cite{MPC}. To illustrate this point, we compare the attainable QoE of the MPC-based ABR that employs a variety of time series prediction techniques against that of the clairvoyant one with access to the true network throughput\footnote{In this work, the term ‘true network throughput’ refers to the inherent transmission rate of the network itself, while the term ‘throughput’ means the measured transmission rate at the client (i.e., an observed value).}. As shown in Fig. 1 (a), an evident QoE gap exists between the clairvoyant and other considered MPC-based ABRs\footnote{The experiment results given in this section employ the HSDPA \cite{HSDPA} network trace, which has the same parameter settings as in Section \uppercase\expandafter{4}.}. This suggests that for video streaming, improving the throughput prediction could result in enhanced user QoE. Besides the observed QoE gap, several other important insights can also be obtained, as detailed in the following.

\textbf{(1) Irregularly Sampled Throughput:} 
% In the application layer, the data are processed in chunks. Thus, historical throughput sequences are collected based on the ratio of the chunk sizes to their transmission latency. Examining the transmission latency of chunks in a streaming session reveals significant fluctuations in the transmission delay. 
In the application layer, past throughputs are irregularly sampled due to the fluctuated transmission delay of the media chunk. However, existing throughput prediction methods use irregular sampling intervals that are equal to the chunk length, ignoring such variability. This can lead to throughput prediction errors, as shown in Fig. 1 (b) where we plot the mean absolute percentage errors (MAPEs) of throughput prediction in training and video streaming. There is a clear increase in MAPE when streaming videos, reflecting the impact of irregular sampling on the throughput prediction accuracy.
\par $\triangleright$ \textbf{Our Solution:} To handle the irregularly sampled throughputs, we utilize a multi-time attention network \cite{mTAN}, which interpolates the irregularly spaced throughput data so that their embeddings become uniformly spaced.
% It is found that using multi-time attention network decreases the throughput prediction errors.

% \textbf{Lack of QoE awareness in throughput prediction:} Existing throughput predictors prioritize prediction accuracy while neglecting the perception of QoE metrics, leading to suboptimal performance of MPC-based ABRs in terms of rebuffering and smoothness, which in turn results in a degradation of QoE.
% We further analyze the significance of uncertainty information and smoothness in throughput prediction, which is given below.\\
\textbf{(2) Prediction Uncertainty:} Due to the ill-posed nature of the time series prediction problem, the prediction techniques inherently have large uncertainty. We give the bitrate and rebuffering results for \textit{trace\_norway\_train\_19.log} in Fig. 2 (a). It can be seen that all methods considered exhibit bitrate overestimation, resulting in a significant rebuffering penalty for QoE and subsequent underestimation of future throughput. This is due to the inaccurate uncertainty information introduced by RobustMPC. Therefore, it is essential to incorporate more precise uncertainty information into throughput prediction to reduce rebuffering events and unexpected underestimation of future throughput.
\par $\triangleright$ \textbf{Our Solution:} To generate reliable uncertainty information for predicted throughput, we design a quantile network \cite{quantile} that provides the distribution of future throughputs. Furthermore, we develop a buffer-aware uncertainty strategy to provide fine-grained uncertainty information, which dynamically adjusts the uncertainty region of the throughput prediction based on buffer occupancy. 
% Empirical results show that our design can produce robust throughput predictions and maintain stable QoE performance. 

\textbf{(3) Smoothness of Throughput Prediction:} During the initial stages of training, we observe that while the prediction accuracy improves, there are only marginal gains in terms of QoE. Analyzing the associated QoE metrics reveals that merely enhancing throughput prediction accuracy may result in significant fluctuation in the predicted values (see Fig. 2 (b) for an illustration), and the lack of prediction smoothness diminishes the advantages brought by improved predictions.
\noindent \par $\triangleright$ \textbf{Our Solution:} To encourage smooth throughput predictions, we include a smoothness regularizer in the loss function when training our throughput predictor. 
\section{System Description}
\par In this section, we describe the design and implementation of Kairos. The workflow of Kairos is shown in Fig. 3. It can be seen that Kairos is composed of two parts, specifically a multi-time attention-based throughput predictor (mTAN-TP) and an uncertainty-aware MPC. They are detailed in Sections \uppercase\expandafter{3}-1 and \uppercase\expandafter{3}-2, respectively.
\subsection{Multi-time Attention-Based Throughput Predictor}
We adopt the multi-time attention network (mTAN) \cite{mTAN} for feature extraction and throughput (as shown in Fig. 4). Unlike other methods that directly process the input sequence, we introduce uniformly spaced reference time points to constrain the temporal positions when generating latent representations. This ensures that these representations obtained via interpolation are uniformly spaced in time, thereby mitigating the impact of irregular sampling on throughput prediction. Besides historical throughput measurements, we also incorporate additional QoE-aware information into mTAN, enabling it to learn a predictor specifically tailored for video streaming. To achieve more flexible sequence prediction based on the uniformly sampled features extracted by mTAN, we design a separate prediction network to compute the quantiles of future throughputs. This allows the prediction sensitivity to adapt to different streaming conditions (see more discussions in Section \uppercase\expandafter{3}-2).

\begin{figure}[t] % 开始图片环境
\centering % 使图片居中显示
\includegraphics[width=0.45\textwidth]{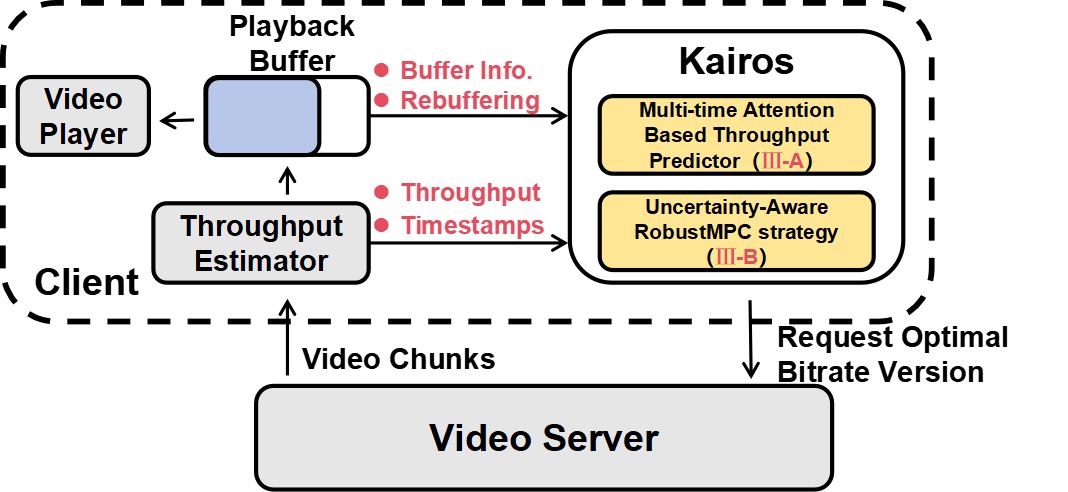} % 插入图片，设置图片宽
% \vspace{-0.3cm}
\caption{Overview of Kairos augmented video streaming system.} % 图片标题
\label{fig:example-image} % 图片标签，用于引用
% \vspace{-0.55cm}
\end{figure}

\par \textbf{Input and Output of mTAN:}
Let $\mathbf{\Phi}_i = \{\boldsymbol{\phi}_{i-k+1}, ..., \boldsymbol{\phi}_i\}$ collect the multi-dimensional historical observations from the past $k$ chunks, where $\boldsymbol{\phi}_i = [{C}_i, B_{i}, T_{i}, D_i]$ is a vector consisting of the throughput, buffer occupancy, rebuffering, and transmission latency for chunk $i$. mTAN takes $\mathbf{\Phi}_i$ and related temporal information as its input $\{\mathbf{\Phi}_{i}, \mathbf{\hat{u}}_{i}, \mathbf{u}_{i}\}$. Here, $\mathbf{\hat{u}}_{i}$ contains $k$ regularly spaced reference time points for interpolation. $\mathbf{u}_{i} = \{t_{i-k+1},...,t_{i}\}$ collects the irregular sampling time instants (i.e., the time instants of download completion for the past $k$ chunks). The latent features extracted by mTAN are subsequently fed to the prediction network, which predicts the throughput of the next chunk $\tilde{C}_{{i+1},q_{[1:M]}}$ at $M$ specified quantile levels $q_{[1:M]}$.

% In order to more effectively grasp the streaming-aware characteristics, we utilize a multi-time attention (mTAN) module to derive hidden features from historical observations. Specifically, After the download of each chunk $i$, mTAN takes the multivariate state $s_i = \{\mathbf{C}_{i}, \mathbf{B}_{i}, \mathbf{T}_{i}, \mathbf{D}_{i}, \mathbf{\hat{u}}_{i}, \mathbf{u}_{i}\}$ into its network. $\mathbf{\hat{u}}_{i}$ represents the past $k$ query time points. This is a uniformly spaced timestamp sequence used to indicate interpolation in the embedding space. $\mathbf{u}_{i}$ represents the starting time of downloads for past $k$ chunks, $\mathbf{C}_{i}$ is the measured throughput for past $k$ chunks, $\mathbf{B}_{i}$ is the buffer occupancy for past $k$ chunks, $\mathbf{T}_{i}$ is the rebuffering time for past $k$ chunks and $\mathbf{D}_{i}$ is the transmission latency for past $k$ chunks. The hidden features extracted by mTAN are subsequently fed into the prediction network $f_{\rm{pred}}$, which predicts next chunk's throughput $\tilde{C}_{i+1}^{q_{[1:M]}}$ at M given quantile levels $q_{[1:M]}$.\\

\textbf{Multi-Time Attention Network:} The mTAN takes $\mathbf{\hat{u}}_i$ as query (Q), $\mathbf{u}_i$ as key (K), and $\mathbf{\Phi}_i$ as value (V). For chunk $i$ (also the current chunk), it returns a latent representation with $k$ elements, which is denoted as $\mathbf{E}_i=\{\mathbf{e}_{i1},...,\mathbf{e}_{ik}\}$. We generate the reference time points $\mathbf{\hat{u}}_i$ using
\begin{equation}
    \mathbf{\hat{u}}_i=\{\hat{u}_n = t_i - n\eta,\ n=1,2,...,k\},
\end{equation}
where $t_i$ is the download completion time for chunk $i$, and $\eta$ is a parameter controlling the distance between two neighboring reference time points. Each element $\mathbf{e}_{in}$ in $\mathbf{E}_i$, $n=1,2,...,k$, has a dimensionality of $J$. For a given query (reference) time point $\hat{u}_{n}$, $\mathbf{e}_{in}$ is found via
\begin{equation}
    \mathbf{e}_{in}[j] = \sum_{h=1}^{H}\sum_{d=1}^{D}\hat{s}_{hd}(\hat{u}_{n},\mathbf{u}_i,\mathbf{\Phi}_i[d]) \cdot P_{hdj},
\end{equation}
where $j=1,2,...,J$. It can be seen that $\mathbf{e}_{in}[j]$ is given by a linear combination of continuous functions $\hat{s}_{hd}(\cdot)$, and $P_{hdj}$ are learnable weights. $\hat{s}_{hd}(\cdot)$ is defined as
\begin{equation}
\hat{s}_{hd}(\hat{u}_{n},\mathbf{u}_i,\mathbf{\Phi}_i[d]) = \sum_{m=0}^{k-1}\kappa_{h}(\hat{u}_{n},\mathbf{u}_i[m])\boldsymbol{\phi}_m[d].
\end{equation}
Mathematically, $\hat{s}_{hd}(\cdot)$ could be considered as a kernelized interpolator in the $d$-th dimension of the observations. The interpolation weights $\kappa_{h}(\cdot)$ are obtained by using an attention mechanism that leverages time embeddings, which is  
\begin{equation}
\label{softmax}
\begin{split}
\kappa_{h}& (\hat{u}_{n}, \mathbf{u}_i[m]) = \\
& \frac{\text{exp}(\alpha_{h}(\hat{u}_{n}) \mathbf{w} \mathbf{v}^T \alpha_{h}(\mathbf{u}_i[m])^T / \sqrt{d_k})}{\sum_{m' = 0}^{k-1} \text{exp}(\alpha_{h}(\hat{u}_{n}) \mathbf{w} \mathbf{v}^T \alpha_{h}(\mathbf{u}_i[m'])^T / \sqrt{d_k})}.
\end{split}
\end{equation}
$\alpha_{h}(\cdot)$ is the time embedding function, which maps input time points into continuous positional embedding. $\mathbf{w}$ and $\mathbf{v}$ are learnable column vectors. The attention mechanism in (\ref{softmax}) is a softmax function. The activation within the softmax is a scaled inner product between the time embedding $\alpha_{h}(\hat{u}_{n})$ of the query time point $\hat{u}_{n}$ and time embedding $\alpha_{h}(\mathbf{u}_i[m])$ of an observed time point $\mathbf{u}_i[m]$, namely the key. Note that $\mathbf{u}_i$ is regenerated as the relative sampling time in the calculation, which is given by $\{t_m=t_i-t_{i-m}, m=0,1,...,k-1 \}$.\\

\textbf{Prediction Module:} After obtaining $\mathbf{E}_i$ using mTAN, we pass it through a gated linear unit (GLU) \cite{GLU}, followed by a skip connection and layer normalization \cite{LN} to facilitate model training. For prediction, the extracted latent feature is fed into a LSTM \cite{LSTM} and additional linear layers to generate quantile prediction outputs $\tilde{C}_{i+1,q}$ for the next chunk $i+1$ at quantile level $q$. The associated computation is given by 
\begin{equation}
   \tilde{C}_{i+1,q[1:M]} = \text{MLP} \left( \text{LSTM} \left( \text{LN} \left( \mathbf{E}_i \odot \sigma(\mathbf{W} \mathbf{E}_i + \mathbf{b}) + \mathbf{E}_i \right) \right) \right).
\end{equation}

\begin{figure}[t]
    \centering
    % \vspace{-0.4cm}
    \includegraphics[width=0.65\linewidth]{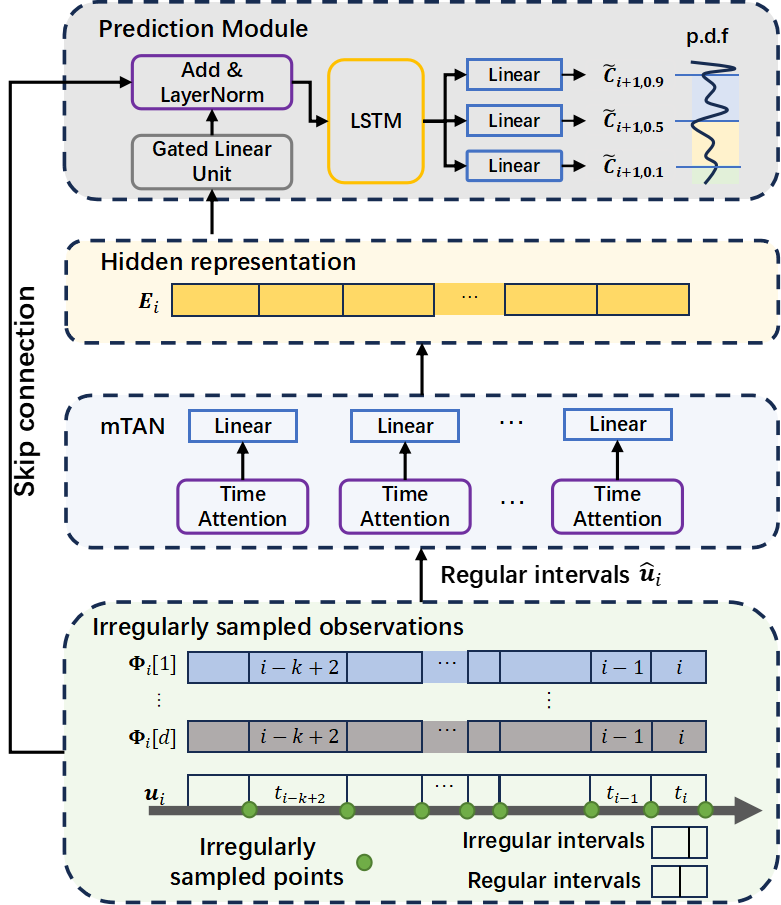}
    % \vspace{-0.2cm}
    \caption{Architecture of the multi-time attention based throughput predictor.}
    \label{fig:enter-label}
    % \vspace{-0.6cm}
\end{figure}

\textbf{Loss Function:} The loss function has two parts: 1) a smoothness regularizer $L_s$ that penalizes the variation in predicted throughputs; and 2) the quantile loss $L_q$ that quantifies the discrepancy between actual throughput and the predicted value. The total loss is thus equal to
\begin{align}
L = L_s+L_q.
\end{align}
\begin{itemize}[leftmargin=*]
\item \textbf{Smoothness Regularizer:} Smoothness is rarely considered in throughput prediction. However, in MPC-based ABR, QoE optimization comes after the throughput prediction is obtained. If two predicted throughputs differ significantly from each other, opting for a smoother bitrate transition might result in a substantial loss in QoE. We thus introduce the following smoothness regularizer that promotes the smoothness of the generated throughput predictions:
\begin{equation}
L_s = 0.1 \cdot L_{sp} + L_{sf},
\end{equation}    
\begin{equation}
L_{sp} = \begin{cases} 
                          \frac{\tilde{C}_{i+1,0.5}}{\frac{\theta}{k} \sum_{n=1}^{k}C_{n}} & \text{if } \frac{\tilde{C}_{i+1,0.5}}{\frac{\theta}{k} \sum_{n=1}^{k}C_{n}} \textgreater 1 \\
                          0 & \text{else.} 
                       \end{cases},
\end{equation}    
\begin{equation}
L_{sf} = \frac{1}{k-2}\sum_{n=3}^{k}(\mathbf{e}_{in } - 2 \mathbf{e}_{i(n-1)} - \mathbf{e}_{i(n-2)})^2.
\end{equation}    

It can be seen that we employ a hierarchical strategy, advocating smoothness both in throughput predictions using $L_{sp}$ and latent representations using $L_{sf}$. The parameter $\theta$ fine-tunes the level of smoothness.
% (in this paper, $\theta$ is 1.1). 
The design of $L_{sp}$ utilizes the principle of additive increase multiplicative decrease (AIMD) \cite{TCP}. It enables the predictor to carry out gradual modifications under stable network conditions while still being able to swiftly adapt to rapid fluctuations in network conditions. More specifically, non-zero smoothness penalty is applied only if the median of the throughput prediction \(\tilde{C}_{i+1,0.5}\) exceeds $\frac{\theta}{k} \sum_{n=1}^{k}C_{n}$.
% insufficient smoothness in the prediction appears, and \(L_{sp}\) is set to \(\frac{\tilde{C}_{i+1,0.5}}{\frac{\theta}{k} \sum_{n=1}^{k}C_{n}}\). Otherwise, \(L_{sp}\) is zero. This allows the model to quickly adapt to decreases in network bandwidth.
% Specifically, we exclusively smooth the median prediction, \(\tilde{C}_{i+1,0.5}\). Specifically, if \(\tilde{C}_{i+1,0.5}\) exceeds \(\theta\) times \(\frac{\tilde{C}_{i+1,0.5}}{\frac{\theta}{k} \sum_{n=1}^{k}C_{n}}\), it indicates insufficient smoothness in the prediction, and \( L_{sp} \) is set to \(\frac{\tilde{C}_{i+1,0.5}}{\frac{\theta}{k} \sum_{n=1}^{k}C_{n}}\). Otherwise, \( L_{sp} \) is zero, allowing the model to quickly adapt to decreases in network bandwidth.
% \par The prediction smoothness $L_{sp}$ aims to make the predictor smoothly rise when the network is stable and quickly drop down when the network is fluctuating. In this paper, we only smooth the median prediction $\tilde{C}_{i+1,0.5}$. Specifically, when $\tilde{C}_{i+1,0.5}$ is larger than $\theta$ times of $\frac{\tilde{C}_{i+1,0.5}}{\frac{\theta}{k} \sum_{n=1}^{k}C_{n}}$, then it indicates that the prediction is not smooth enough and $L_{sp}$ equals $\frac{\tilde{C}_{i+1,0.5}}{\frac{\theta}{k} \sum_{n=1}^{k}C_{n}}$. Otherwise, $L_{sp}$ equals 0 so that the model can react quickly to network bandwidth dropping.
$L_{sf}$ intends to smooth the interpolated throughput results produced by mTAN. Similar to \cite{Smoother}, we utilize a second-order regularizer to enhance the smoothness of the extracted latent mTAN features.
\item \textbf{Quantile Loss Function \cite{quantile}:}
The quantile loss function $L_q$ measures the prediction errors across different quantiles, which is defined as
\begin{equation}
\begin{aligned}
    L_{q} = \frac{1}{N}\sum_{i=1}^{N} \sum_{q \in q[1:M]} & q \cdot \text{max}(0, \tilde{C}_{i,q} - \hat{C_i}) \\
& + (1 - q) \cdot \text{max}(0, \hat{C_i} - \tilde{C}_{i,q}). 
\end{aligned} 
\end{equation}
$\tilde{C}_{i,q}$ denotes the prediction under quantile $q$, $\hat{C}_i$ is the ground truth. Specifically, when $q = 0.5$, the quantile loss is simply the mean absolute error (MAE). % We select different values for $q$ to generate prediction outputs for different quantiles.
\end{itemize}

\begin{figure*}[ht]
\centering
\vspace{-0.3cm}
\begin{subfigure}[b]{0.22\textwidth}
\includegraphics[width=\textwidth]{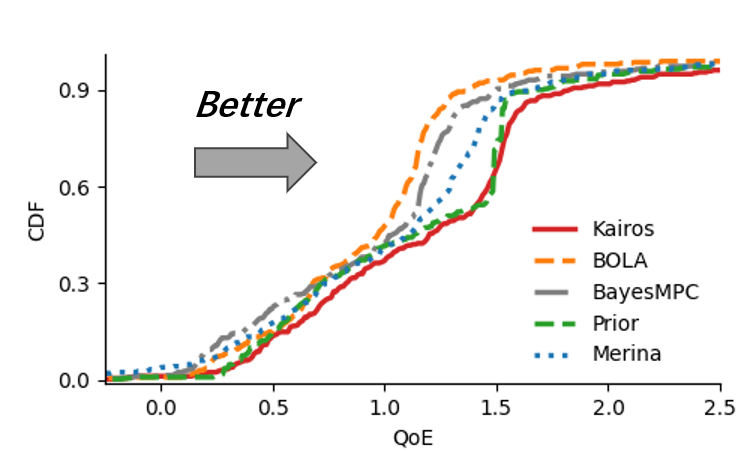}
% \caption*{} % 去除标题
% \vspace{-0.5cm}
\caption{HSDPA+FCC traces}
% \label{fig:sub1}
\end{subfigure}
\hspace{2mm} % Adjusts the spacing between images
% \hfill
\begin{subfigure}[b]{0.22\textwidth}
\includegraphics[width=\textwidth]{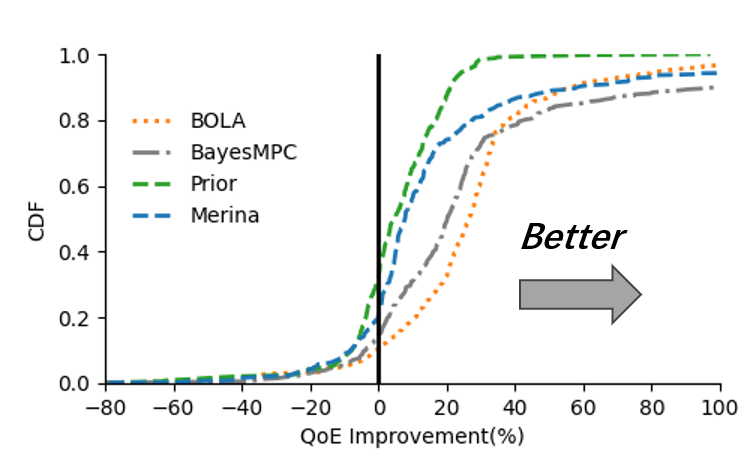}
% \caption*{} % 去除标题
% \vspace{-0.5cm}
\caption{HSDPA+FCC traces}
% \label{fig:sub1}
\end{subfigure}
\hspace{2mm} % Adjusts the spacing between images
% \hfill
\begin{subfigure}[b]{0.22\textwidth}
\includegraphics[width=\textwidth]{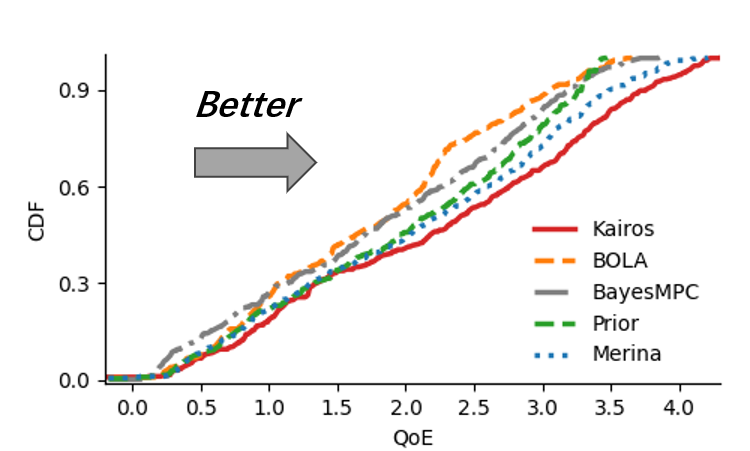}
% \caption*{} % 去除标题
% \vspace{-0.5cm}
\caption{OBOE traces}
% \label{fig:sub2}
\end{subfigure}
\hspace{2mm} % Adjusts the spacing between images
% \hfill
\begin{subfigure}[b]{0.22\textwidth}
\includegraphics[width=\textwidth]{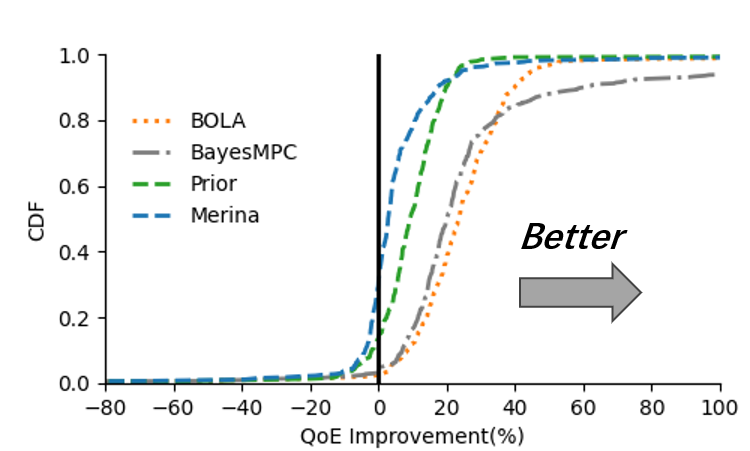}
% \caption*{} % 去除标题
% \vspace{-0.5cm}
\caption{OBOE traces}
% \label{fig:sub2}
\end{subfigure}
% \vspace{-4mm} % Adjusts the spacing between images
\caption{Comparing QoE performance of Kairos with existing ABR algorithms on all considered network traces.}
\label{fig:cdf-hsdpa} % Image label for referencing
 % \vspace{-0.4cm}
\end{figure*}

\subsection{Uncertainty-Aware MPC}
\textbf{Buffer-aware Uncertainty Adjustment:} Throughput prediction using quantiles inherently incorporates uncertainty, thus enhancing fault tolerance. Nevertheless, choosing an appropriate quantile level is challenging, as a fixed quantile may lead to strategies that are either conservative or aggressive, which results in excessively optimistic or pessimistic predictions. To mitigate this issue, we combine the quantile for throughput prediction with buffer occupancy, allowing for dynamic adjustment of the prediction sensitivity while accounting for buffer variations. Mathematically, we have that 
\begin{equation}
    \gamma = \alpha + \beta / B_i,
\end{equation}
\begin{equation}
\label{Predicted throughput}
    \Tilde{C_i}= \Tilde{C}_{i,0.5} - \gamma \cdot (\Tilde{C}_{i,0.5} - \Tilde{C}_{i,0.1}).
\end{equation}
% where $\alpha$ and $\beta$ are fitting parameters.
% (in this paper, $\alpha$ = 0.75, $\beta$ = 1.2), 
In this work, the difference between the 0.1 quantile and 0.5 quantile predictions is used to define the uncertainty interval. The parameter $\gamma$ serves as a tuning factor that dynamically scales the uncertainty interval. The value $\Tilde{C_i}$ represents the adjusted throughput prediction. When the size of the buffered content is large, it encourages more aggressive bitrate decisions, as the presence of buffered content can accommodate network variability. When the remaining buffered content is small, predictions are down-scaled to promote more conservative bitrate decisions, reducing the chance of rebuffering.
\par \textbf{MPC Strategy with Uncertainty Information:} We apply the throughput prediction $\Tilde{C}_i$ in Eq. (\ref{Predicted throughput}) to MPC to generate a robust bitrate decision policy. The bitrate decision strategy is formulated as the following maximization problem:
\begin{align}
&\mathop{\max}\limits_{R_m,\cdots,R_{m+N-1}} \sum_{i=m}^{m+N-1} \text{QoE}_i \\
& \text{QoE}_i = R_i - \lambda T_i -\mu |R_i-R_{i-1}|\\
\text{s.t.} \quad 
&\begin{cases}
T_i = \max\left(0,\frac{R_i \hat{D}_i}{\Tilde{C}_i} - B_i\right) \\
B_{i+1} = B_i + \hat{D}_i - \frac{R_i \hat{D}_i}{\Tilde{C}_i} \\
t_{i+1} = t_i + \frac{R_i \hat{D}_i}{\Tilde{C}_i}\\
R_i \in \{r_1, r_2, \cdots, r_L\}.
\end{cases}
\vspace{-0.2cm}
\end{align}
where \( r_i \) represents the available bitrate levels, $\hat{D}_i$ is the chunk duration, $T_i$ denotes rebuffering time, the buffered content size $B_i\in[0,B_{\text{max}}]$, $\lambda$ and $\mu$ are penalty factors for rebuffering and bitrate smoothness. The MPC algorithm utilizes the past \( k \) observations to solve the above optimization problem based on throughput predictions. It aims to maximize the QoE for \( N \) future chunks and determine the optimal bitrate sequence \( R_{[m, m+N-1]}^{*} \). Subsequently, \( R_{m}^{*} \) is used as the optimal bitrate to request the next chunk. 

\section{Experiments}
In this section, we extensively evaluate the performance of Kairos against existing ABR algorithms under various network conditions.

% \textbf{Outline:} This section includes:
\subsection{Experimental Setups}
% \textbf{Testbed Setups:} Both trace-driven and real-world experiments are conducted to evaluate the performance of Kairos.
% \begin{itemize}[leftmargin=*]
% \item[$\triangleright$] \par \textbf{Trace-driven Emulation:}
% In trace-driven emulation, we test ABR algorithms on the virtual player provided by Pensieve \cite{PENSIEVE}. The virtual player simulates HTTP video streaming using network traces collected in real world. For more details on virtual player setup, readers are referred to \cite{PENSIEVE}.
% \item[$\triangleright$] \par \textbf{Real-world Deployment:}
% To further examine the design of Kairos, we create a real-world testbed. An Ubuntu-based server is set up to host several video representations and store the \textit{.XML} file for chunk description. Another desktop acts as the client and connects with the server via real-world networks. It runs Kairos to fetch video chunks from the server and playback using the \textit{dash.js} player \cite{Dash}.
% \end{itemize}

\par \textbf{Video Datasets:} We use Envivio Dash3 \cite{Envivio} for streaming. The video is encoded using the H.264 codec \cite{H264} with a bitrate ladder of \{0.3, 0.75, 1.2, 1.85, 2.85, 4.3\} Mbps. The video is 192 seconds long and divided into 48 chunks, each lasting for 4 seconds.
% The video has a total duration of 192 seconds and is segmented into 48 chunks, each lasting for 4 seconds.
\par \textbf{Network Traces Datasets:} We create a corpus of traces from existing real network datasets, including 3G/HSDPA \cite{HSDPA}, FCC \cite{FCC}, OBOE \cite{OBOE} and FUGU\cite{Fugu}. The throughput of these traces varies from 0.11 to 12.85 Mbps, aligned with our video bitrate range. We leverage FUGU traces for model training and the remaining datasets for evaluation.
% \par \textbf{Model Training:} We leverage FUGU real-world video streaming logs \cite{Fugu} to generate more than 100,000 sequence samples for model training. 
% We gather 1000 real-world data logs from FUGU and generate 100,000 samples from them. 80\% of these samples are used for training and the remaining 20\% are for validation. We set the training epoch to 100, and the process runs on a desktop equipped with an i7-7700K CPU with 8 cores and a GTX 1080 Ti GPU. The entire training completes in just 0.1 hours.
\par \textbf{QoE Metrics:} As the QoE preference is not our focus, we employ the linear QoE model, which is a linear combination of video quality, rebuffering time, and bitrate smoothness. The average QoE is defined as $\text{QoE} = \frac{1}{N-1} \sum_{i=2}^{N}\text{QoE}_i$, where $N$ denotes the total number of video chunks, and $\text{QoE}_i$ is computed using Eq. (14). For consistency with prior works, we use $\lambda=4.3$ and $\mu=1.0$. 
\par \textbf{ABR Baselines:} 
\textbf{(1) BOLA \cite{BOLA}:} a buffer-based ABR that uses the Lyapunov optimization to select optimal bitrate for the next video chunk with buffer occupancy constraints. 
\textbf{(2) BayesMPC \cite{BayesMPC}:} a state-of-art MPC-based ABR streaming method that leverages Bayesian neural networks to predict the lower bound of future throughputs and make ABR decisions. 
\textbf{(3) Prior \cite{Prior}:} a learning-based ABR method that combines attention-based throughput predictor with pensieve-ppo \cite{PENSIEVE} to enhance ABR decisions. 
\textbf{(4) Merina \cite{MERINA}:} a learning-based ABR streaming method that utilizes meta learning and deep reinforcement learning to select bitrate for the next video chunks.

\begin{figure}[t]
\centering
% \vspace{-0.1cm}
\begin{subfigure}[b]{0.5\textwidth}
\includegraphics[width=\textwidth]{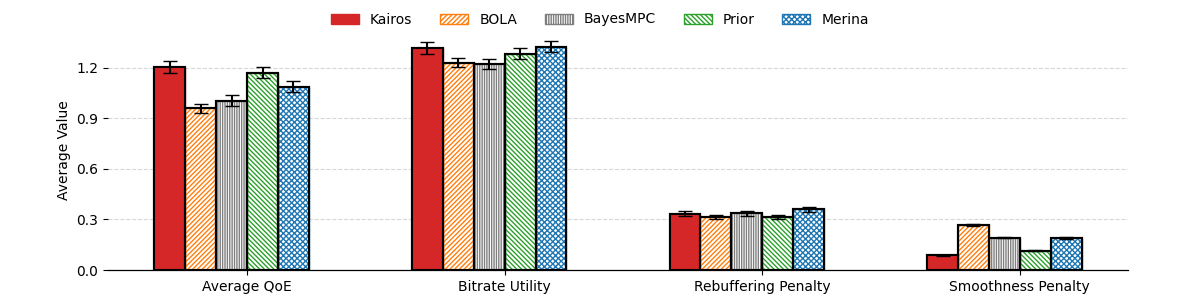}
% \caption*{} % 去除标题
% \vspace{-0.5cm}
\caption{FCC+HSDPA traces}
\label{fig:sub1}
\end{subfigure}
\hspace{1mm} % Adjusts the spacing between images
% \hfill
\begin{subfigure}[b]{0.5\textwidth}
\includegraphics[width=\textwidth]{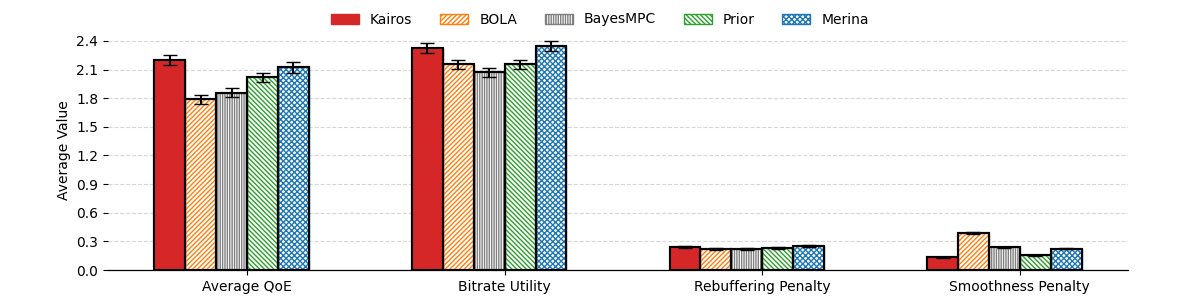}
% \caption*{} % 去除标题
% \vspace{-0.5cm}
\caption{OBOE traces}
\label{fig:sub1}
\end{subfigure}
\vspace{-0.4cm}
\caption{Comparing QoE submetrics of Kairos with existing ABR algorithms on all considered network traces.}
\label{fig:submetircs}
% \vspace{-0.55cm}
\end{figure}

\subsection{Kairos versus Other ABRs}
\par In this experiment, we use trace-driven emulation to compare the QoE performance of Kairos against that of the ABR baselines on HSDPA, FCC and Oboe traces.
\par Fig. \ref{fig:cdf-hsdpa} shows the QoE performance on all network datasets. We observe that Kairos outperforms the second best baseline (Prior in FCC+HSDPA traces and Merina in OBOE traces), by 6.42\% to 6.83\% increase in the average QoE with a reduction of 55.07\% to 64.17\% in smoothness penalty. We also record the QoE improvement in percentage that Kairos achieves over other ABR schemes. The findings indicate that Kairos surpasses ABR baselines in more than 67.68\% of sessions. Furthermore, we report the performance of key metrics including bitrate utility, rebuffering penalty, smoothness penalty, with average QoE as shown in Fig.\ref{fig:submetircs}. The results demonstrate that while Kairos may not deliver the top performance in every QoE submetric, it achieves the highest overall QoE across all considered network conditions. Merina performs similarly to Kairos in terms of bitrate utility and rebuffering but falls short in smoothness as it does not adapt well to unknown environments. Prior and BayesMPC exhibit suboptimal performance in bitrate utility due to their conservative manner. BOLA fails to balance QoE submetrics and thus performs poorly on overall QoE. 
% Although FESTIVE and RobustMPC achieve better bitrate utility, they struggle with rebuffering and fail to provide smooth bitrate transitions due to overly aggressive throughput predictions.

\begin{figure}[t]
% \vspace{-0.1cm}
\centering
  \begin{subfigure}[b]{0.22\textwidth}
    \includegraphics[width=\textwidth]{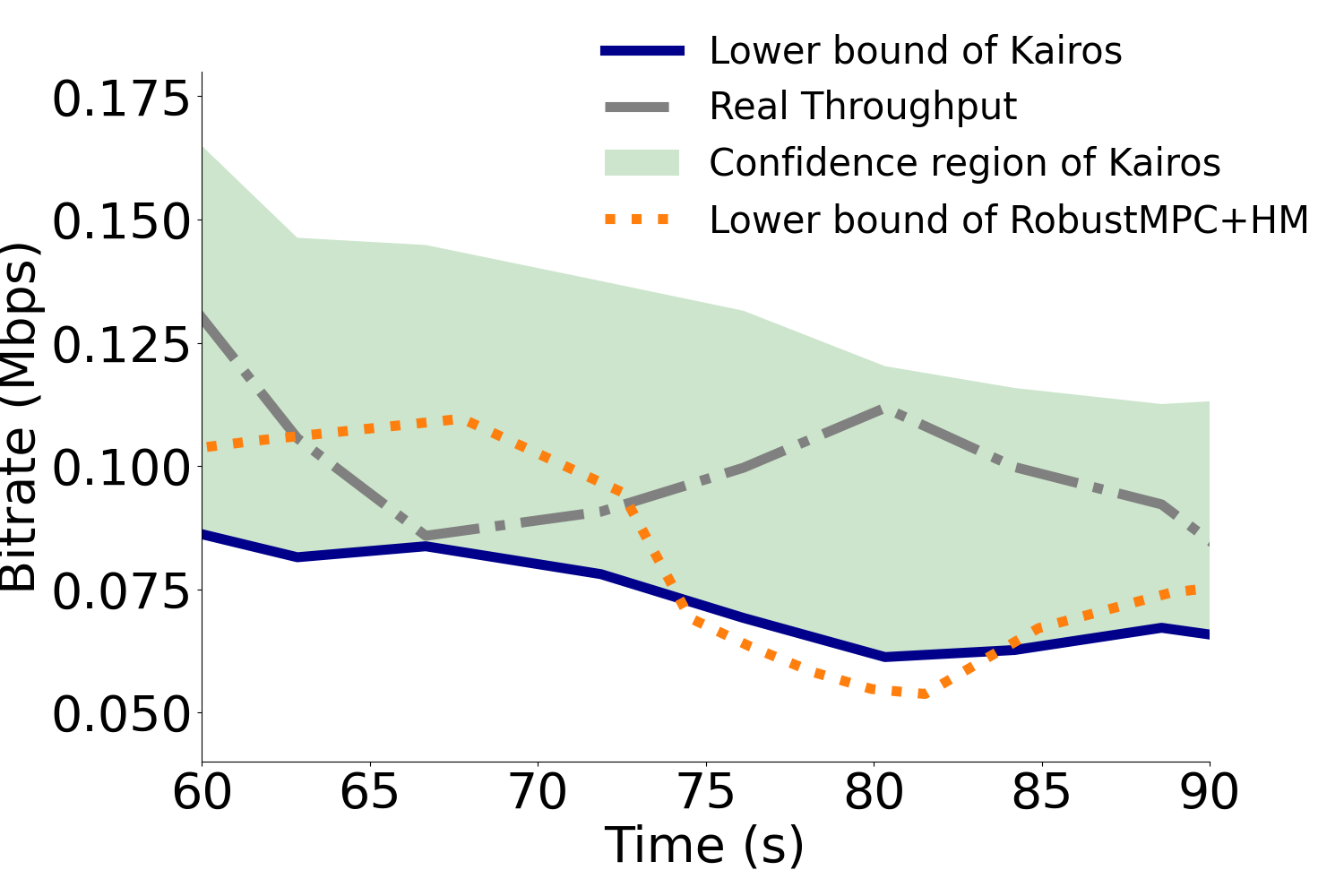}
    \vspace{-0.6cm}
    \caption{trace\_norway\_bus\_13.log}
    \label{fig:7a}
  \end{subfigure}
  \hspace{3mm} % Adjusts the spacing between images
  \begin{subfigure}[b]{0.22\textwidth}
    \includegraphics[width=\textwidth]{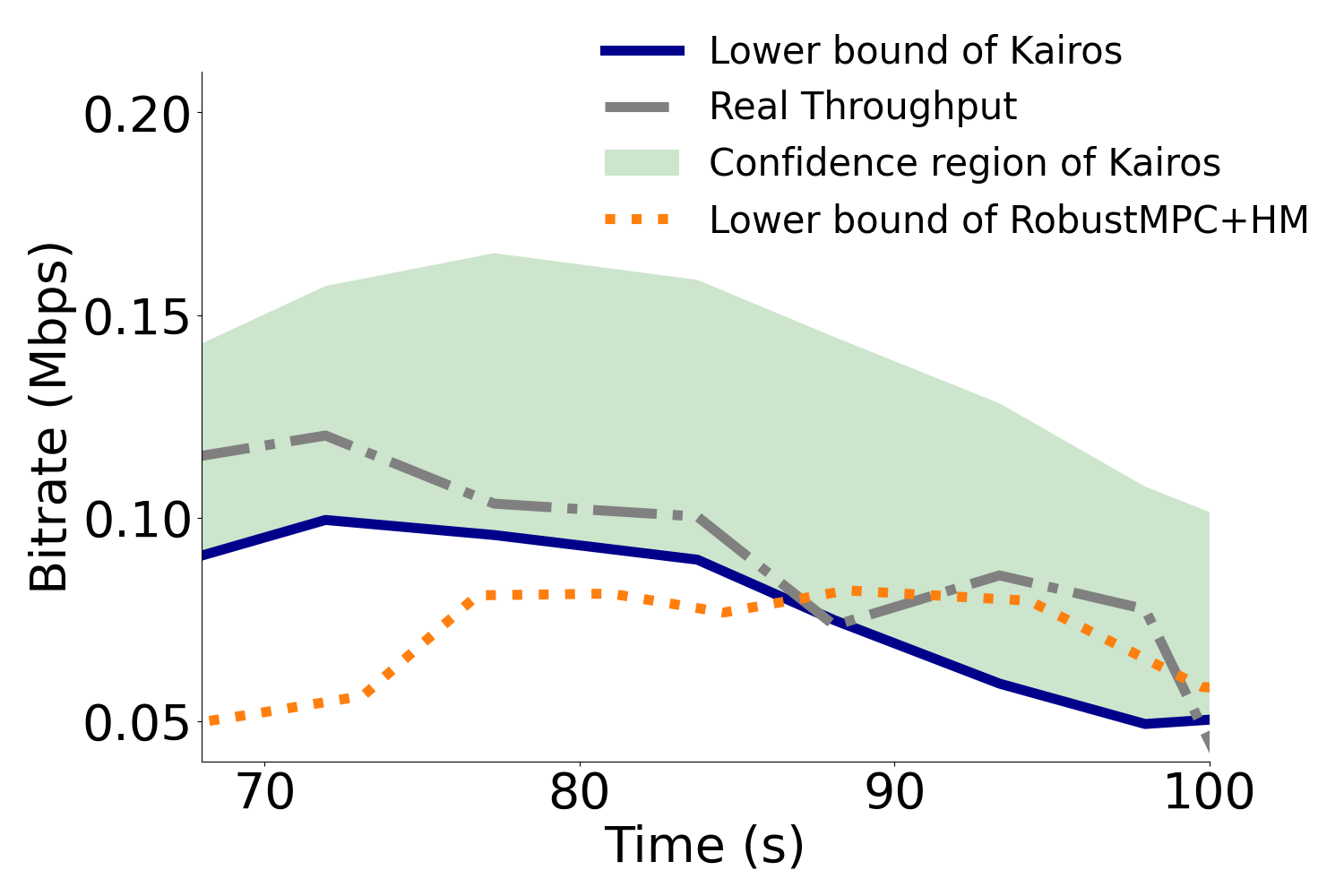}
    \vspace{-0.6cm}
    \caption{trace\_norway\_tram\_12.log}
    \label{fig:7b}
  \end{subfigure}
% \vspace{-0.4cm}
\caption{A case study of Kairos and RobustMPC+HM.}
\label{fig:case-study} % Image label for referencing
% \vspace{-0.6cm}
\end{figure}

\textbf{Case Study:} We randomly select two specific traces to illustrate Kairos's behavior. As shown in Fig. \ref{fig:case-study}, we compare Kairos with vanilla RobustMPC. The results demonstrate that Kairos effectively predicts the lower bound of future throughput that matches the true network throughput and reacts swiftly to throughput changes compared to vanilla RobustMPC, thereby leading to higher QoE performance.

\subsection{Evaluations of Throughput Predictors}
In this study, we evaluate the accuracy of various throughput predictors by analyzing the log files gathered from the FCC, HSDPA, and FUGU real network datasets. 
% We choose several representative throughput prediction algorithms for this purpose, including:
% \begin{itemize}[leftmargin=*]
%     \item[$\triangleright$] HM: It predicts future throughput by calculating the harmonic mean of several past throughput samples.
%     \item[$\triangleright$] LSTM: Mei et al. \cite{LSTMPred} proposes using an LSTM model to predict real-time mobile throughput based on historical samples.
%     \item[$\triangleright$] DTP: Huang \cite{DeepMPC} proposes using 1D-CNN to predict the next chunk's throughput based on past samples.
%     \item[$\triangleright$] mTAN-TP: The throughput predictor used in Kairos that captures the irregularities inherent in the input sequence to attain better prediction accuracy.
% \end{itemize}

\par
\begin{figure}[t]
    % \vspace{-0.3cm}
    \centering
    \includegraphics[width=0.8\linewidth]{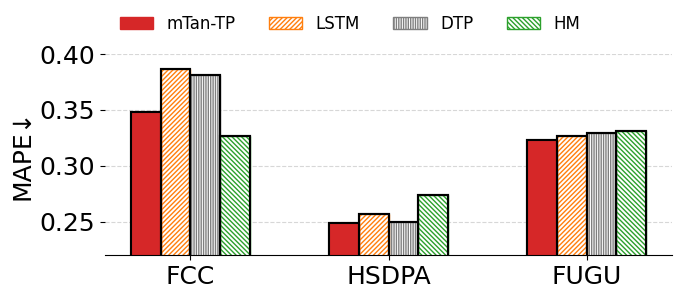}
    % \vspace{-0.45cm}
    \caption{MAPE of throughput predictors over different network traces.}
    \label{fig:mape}
    % \vspace{-0.35cm}
\end{figure}

\begin{figure}[t]
\centering
% \vspace{-0.2cm}
  \begin{subfigure}[b]{0.22\textwidth}
    \includegraphics[width=\textwidth]{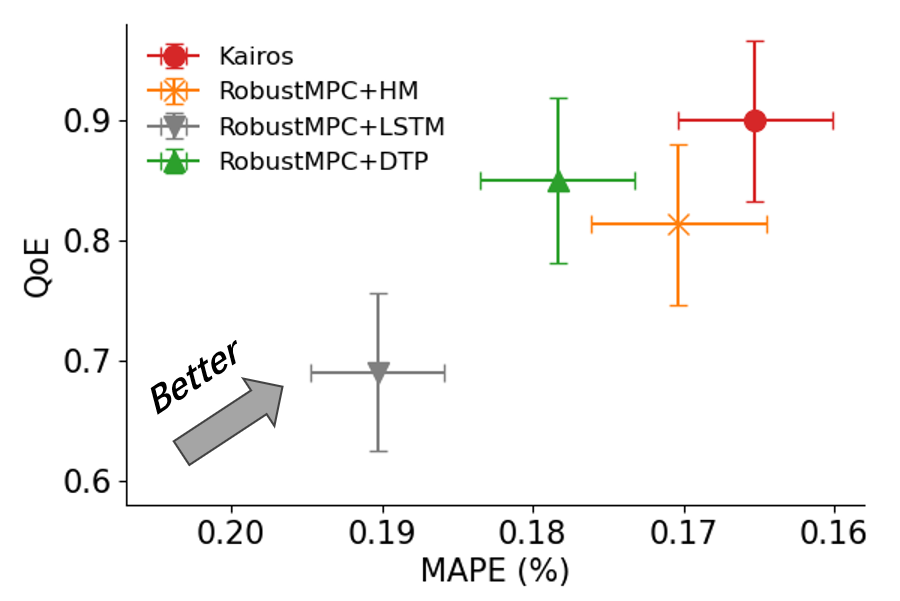}
    % \vspace{-0.6cm}
    \caption{qoe vs. prediction error}
    % \vspace{-0.4cm}
    \label{fig:sub1}
  \end{subfigure}
  \hspace{2mm} % Adjusts the spacing between images
  \begin{subfigure}[b]{0.22\textwidth}
    \includegraphics[width=\textwidth]{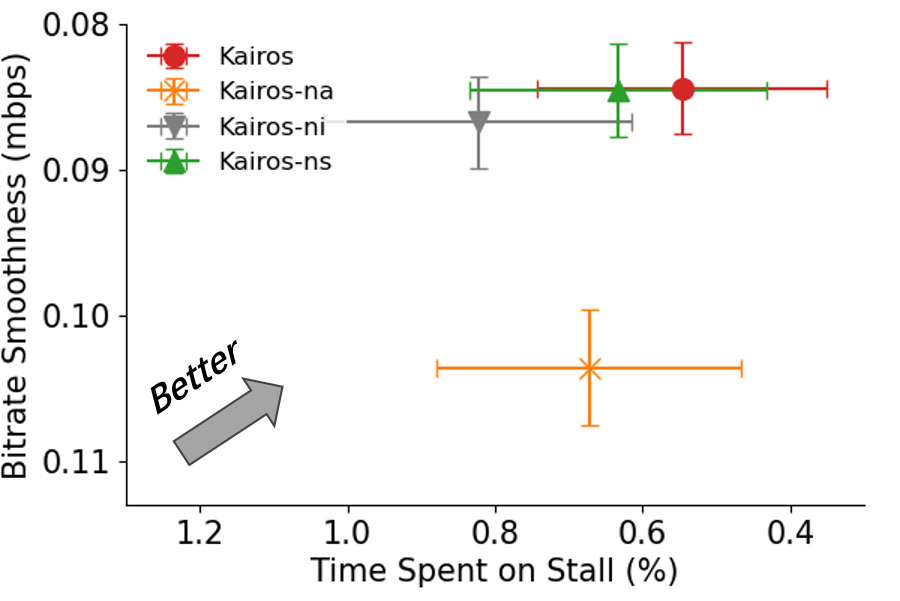}
    % \vspace{-0.6cm}
    \caption{smoothness vs. rebuffering}
    % \vspace{-0.4cm}
    \label{fig:sub1}
  \end{subfigure}
  
  % \hspace{3mm} % Adjusts the spacing between images
\caption{Comparing Kairos with its variants and other throughput predictor assisted MPC-based ABR schemes on HSDPA and FCC datasets.}
\label{fig:abl} % Image label for referencing
% \vspace{-0.4cm}
\end{figure}

 As shown in Fig. \ref{fig:mape}, the rule-based method HM \cite{MPC} relying solely on the harmonic means of the past observations, struggles to predict future throughput accurately under fluctuating network conditions. Despite HM outperforming other learning-based predictors under FCC datasets, it cannot guarantee satisfactory QoE performance (As shown in Fig. \ref{fig:abl} (a)). This is due to the fact that high prediction accuracy is susceptible to overestimated throughputs when an unpredictable throughput drop is present, potentially triggering severe rebuffering events and a subsequent decrease in QoE. Unlike HM, learning-based algorithms leverage data-driven methods to predict future throughput. For instance, LSTM \cite{LSTM} utilizes a recurrent network to capture temporal features while DTP \cite{DeepMPC} uses 1D-CNN to capture the hidden features of network dynamics. Nonetheless, both fail to capture the irregular sampling nature of historical observations, which causes the performance gap in video streaming as mentioned in Fig. 1. mTAN-TP considers the irregularities of the historical series to enhance the extraction of the intrinsic features within the input sequences, thereby achieving superior predictive accuracy.

% \begin{table}[t]
% \centering
% \caption{MAPE of existing throughput prediction algorithms over different network traces.}
% \label{tab:mape}
% \begin{tabular}{cccc} % 定义了三列
% \toprule
% & FCC & HSDPA & FUGU \\
% \midrule
% HM & \textbf{0.327} & 0.274 & 0.331\\
% LSTM & 0.387 & 0.257 & 0.327 \\
% DTP & 0.381 & 0.250 & 0.329\\
% \rowcolor{gray!20}
% \textbf{mTAN-TP} & 0.348 & \textbf{0.249} & \textbf{0.323}\\
% \bottomrule
% \end{tabular}
% \end{table}

\subsection{Ablation Study}
\par In this section, we investigate the credibility of each component of Kairos to gain a clearer insight into the design philosophy behind Kairos. We develop three variants of Kairos: \textbf{(1) Kairos-na:} it replaces the uncertainty adjustment module with the discount factor of RobustMPC; \textbf{(2) Kairos-ni:} it removes mTAN and retains the LSTM for throughput prediction; \textbf{(3) Kairos-ns:}  it removes the smoothness regularizer during throughput predictor's training. 
\par We compare the rebuffering and smoothness performance of Kairos and its variants over HSDPA and FCC datasets. As shown in Fig. \ref{fig:abl} (b), thanks to the smoothness regularizer and the uncertainty adjustment module, Kairos significantly lowers rebuffering and smoothness penalties, leading to more accurate throughput prediction and decreased overestimation. Besides, we notice that even though Kairos-ns removes the smoothness regularizer during training, it still obtains a relatively low smoothness penalty. This is due to the precisely predicted lower bound of the future throughput can help MPC attain better smoothness and thus the smooth regularizer serves as a compensating factor to further enhance the bitrate smoothness.

\subsection{Real-world Experiment}
In this experiment, we compare Kairos with two representative ABR schemes (i.e. RobustMPC \cite{MPC} and Pensieve \cite{PENSIEVE}) under several types of real world connections. 
\par
\begin{figure}[t]
\centering
% \vspace{-0.3cm}
    \begin{subfigure}[b]{0.22\textwidth}
    \includegraphics[width=\textwidth]{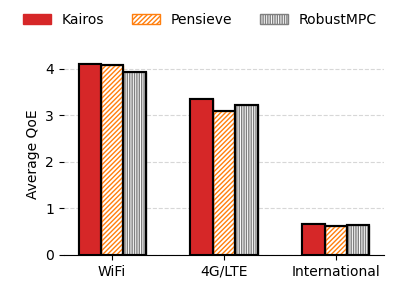}
       % \vspace{-0.6cm}
    % \caption{trace\_norway\_bus\_13.log}
    \label{}
    \end{subfigure}
    \hspace{2mm} % Adjusts the spacing between images
    \begin{subfigure}[b]{0.22\textwidth}
    \includegraphics[width=\textwidth]{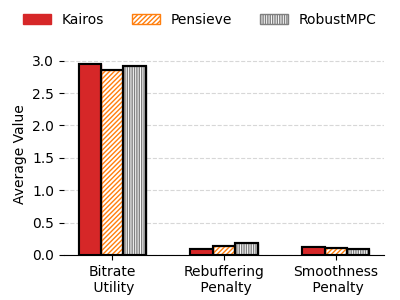}
    % \caption{trace\_norway\_tram\_12.log}
      % \vspace{-0.6cm}
    \label{}
    \end{subfigure}
\vspace{-0.2cm}
\caption{Comparing Kairos with ABRs in real world.}
\label{fig:real-world} % Image label for referencing
% \vspace{-0.5cm}
\end{figure}

 In our evaluation, the performance tests are conducted on different network conditions, including campus Wi-Fi, 4G/LTE cellular networks, and international networks. The international networks are evaluated from Shanghai, China, to San Jose, USA.
% The international network tests involved two routes: one from Shanghai, China, to San Jose, USA, and the other from Shanghai, China, to Christchurch, New Zealand. 
% \begin{table}[t]
% \centering
% \caption{Real-world network metrics.}
% \label{tab:network_metrics}
% \begin{tabular}{cccc}
% \toprule
% Network & RTT (ms) & $\mu$ (Mbps) & $\sigma$ \\
% \midrule
% WiFi & 32.62 & 13.68 &  10.88 \\
% 4G/LTE & 34.97 & 7.12 & 8.48 \\
% International & 99.72 & 1.52 & 1.44 \\
% \bottomrule
% \end{tabular}
% \end{table}
Specifically, for each session, we randomly pick up one ABR scheme, and the same test video is played in full. For each type of network connection, a 1-hour testing period is performed for the algorithm. The results are reported with the average QoE of each session. As shown in Fig. \ref{fig:real-world}, Kairos outperforms comparable or better than Pensieve and RobustMPC under all considered networks. Additionally, Kairos achieves enhancements in average QoE by 0.48\% to 8.43\%. The improved performance stems from Kairos' ability to swiftly adapt to changes in network conditions. This highlights the importance of incorporating a streaming-aware throughput predictor in MPC-based ABR for achieving superior QoE.

% \textbf{Baselines:} 
%  \begin{itemize}
%      \item (r)MPC
%      \item BOLA
%      \item FESTIVE (Rate-based)
%      \item Pensieve
%      \item BayesianMPC
%      \item (MPC+LSTM)/Fugu?
%  \end{itemize}
% \textbf{Traces:}
%  FCC; HSDPA; Oboe; Fugu \\
% \textbf{QoE metrics:} Linear model

% \begin{table}[t]
% \vspace{-0.4cm}
% \centering
% \caption{Real-world network metrics.}
% \vspace{-0.4cm}
% \label{tab:network_metrics}
% \begin{tabular}{cccc}
% \toprule
% Network & RTT (ms) & $\mu$ (Mbps) & $\sigma$ \\
% \midrule
% WiFi & 32.62 & 13.68 &  10.88 \\
% 4G/LTE & 34.97 & 7.12 & 8.48 \\
% International & 99.72 & 1.52 & 1.44 \\
% \bottomrule
% \end{tabular}
% \end{table}

% \subsection{Performance comparison of ABRs}
% \begin{itemize}
%     \item Compare our scheme with baselines over all traces
%     \item QoEs CDF
%     \item Submetric (bitrate, rebuffering, smoothness) bar diagram
% \end{itemize}
% \subsection{Our bandwidth prediction scheme with other ABRs}
% \begin{itemize}
%     \item RB; HYB
%     \item QoE improvement CDF
%     \item submetrics
% \end{itemize}

% \subsection{Throughput comparison}
% \begin{itemize}
%     \item LSTM, HM
%     \item MSE. MAP
% \end{itemize}

% \subsection{Ablation study}
% \begin{itemize}
%     \item Incremental ablation study
%     \item w/o. prediction rmpc
%     \item w/o. smoothness mtan-ns
%     \item w/o. discounting factor mtan + robust-discount
%      accuracy 比较预测器特性
%      Kairos、Kairos-No-smooth、Kairos-No-adaptive、Kairos-No-irregular、 对比不同指标，说明每个设计的重要性，对比rebuferring和smoothness
%      mTan、CNN、GCA、LSTM、harmonic mean、对比QoE和accuracy箱线图,
%     \item qoe cdf overall comp
% \end{itemize}

% \subsection{Real-world experiment}
% \begin{itemize}
%     \item WiFi; wired(constrained); 4G cellular
%     \item QoE + submetrics
%     \item QoE bar 
% \end{itemize}

% \section{Discussion}
% \textbf{Outline:} This section discusses the significance of our research and the design philosophy.
% tatp
% \begin{itemize}
%     \item Discuss the importance of uniform throughput prediction
%     \item Discuss the rationale of smoothness
%     \item Discuss training and inference overhead 
% \end{itemize}

\section{Related Work}
Adaptive bitrate (ABR) algorithm has been extensively researched in recent years (\cite{FESTIVE, BBA, BOLA, GENET, Karma, MetaABR}). Most ABR systems rely on throughput predictions to make bitrate decisions. The earliest throughput estimation techniques utilize the harmonic mean of the past observations to forecast future throughput. CS2P \cite{CS2P} models the evolving throughput of users as a hidden Markov chain with a limited number of discrete states. Lumos \cite{Lumos} examines the influence of application behavior and proposes a decision tree-based throughput predictor. Recent studies (\cite{LSTMPred, Prior}) also employ deep learning models (\cite{LSTM, Transformer}) to achieve more accurate throughput predictions. All the aforementioned approaches use deterministic methods for throughput prediction, which can perform poorly in certain scenarios. By considering uncertainty, BayesMPC \cite{BayesMPC} leverages BNN to predict the probability distribution of future throughputs based on historical observations. However, these methods overlook the irregular sampling characteristics in historical data and the importance of smooth prediction in video streaming, resulting in suboptimal QoE metrics. 
% \par In contrast to previous research, our work focuses on designing a streaming-aware throughput predictor to improve QoE performance for video streaming. By handling irregularly sampled issues and designing a uncertainty prediction strategy with a smoothness regularizer, our scheme achieves satisfied QoE performance under various network conditions.
% \textcolor{red}{(this part seems incomplete right now (the last sentence should not only include their drawbacks, but also a brief intro about our modification), shorten Section 2 to make this section more complete)}

\section{Conclusion}
In this paper, we present Kairos, a novel MPC-based ABR scheme that integrates a streaming-aware throughput predictor for video streaming. Kairos addresses the irregularities inherent in input sequences by employing a multi-time attention network, ensuring more accurate modeling of temporal dynamics. Furthermore, it incorporates a dedicated prediction network that estimates the distribution of future throughput and utilizes a buffer-aware uncertainty adjustment module to establish a conservative lower bound, thereby aiding MPC in making optimal bitrate decisions. Additionally, we embed a smoothness regularizer within the throughput predictor to mitigate abrupt bitrate changes. Our comprehensive experiments demonstrate that Kairos significantly outperforms existing state-of-the-art ABR methods under diverse real-world network conditions, offering robust and smooth performance improvements.

% \begin{acks}
% This paper is supported in part by National Natural Science Foundation of China (62371290), National Key R\&D Program of China (2024YFB2907204), the Fundamental Research Funds for the Central Universities of China, and STCSM under Grant (22DZ2229005). The corresponding author is Yiling Xu(e-mail: yl.xu@sjtu.edu.cn).
% \end{acks}

\bibliographystyle{ACM-Reference-Format}
\bibliography{sample-sigconf-camera-ready}

\end{document}